\documentclass[prd,superscriptaddress, nofootinbib,showpacs,preprintnumbers,floatfix]{revtex4}

\usepackage{mathrsfs}
\usepackage{amsfonts,amssymb,amsmath}
\usepackage{graphicx,epsfig}
\usepackage{multirow}

\def\fsu5{$\cal{F}$-$SU(5)$}

\def\m1half{$M_{1/2}$}
\def\m3half{$M_{3/2}$}
\def\m32{$M_{32}$}

\begin{document}

\title{No-Scale Multiverse Blueprints at the LHC}

\author{Tianjun Li}

\affiliation{ State Key Laboratory of Theoretical Physics, Institute of Theoretical Physics,
Chinese Academy of Sciences, Beijing 100190, P. R. China }

\affiliation{George P. and Cynthia W. Mitchell Institute for Fundamental Physics and Astronomy,
Texas A$\&$M University, College Station, TX 77843, USA }

\author{James A. Maxin}

\affiliation{George P. and Cynthia W. Mitchell Institute for Fundamental Physics and Astronomy,
Texas A$\&$M University, College Station, TX 77843, USA }

\author{Dimitri V. Nanopoulos}

\thanks{\textit{\textbf{\normalsize Contribution to the Proceedings of the International Symposium on "Subnuclear Physics: Past, Present and Future" held at the Pontifical Academy of Sciences, Vatican City, 30 October - 2 November 2011, based on a talk given by
Dimitri V. Nanopoulos.}}}

\affiliation{George P. and Cynthia W. Mitchell Institute for Fundamental Physics and Astronomy,
Texas A$\&$M University, College Station, TX 77843, USA }

\affiliation{Astroparticle Physics Group, Houston Advanced Research Center (HARC),
Mitchell Campus, Woodlands, TX 77381, USA}

\affiliation{Academy of Athens, Division of Natural Sciences,
28 Panepistimiou Avenue, Athens 10679, Greece }

\author{Joel W. Walker}

\affiliation{Department of Physics, Sam Houston State University,
Huntsville, TX 77341, USA }


\begin{abstract}

We review the blueprints of the No-Scale multiverse and the LHC search.

\end{abstract}

\pacs{11.10.Kk, 11.25.Mj, 11.25.-w, 12.60.Jv}

\preprint{ACT-02-12, MIFPA-12-04}

\maketitle


\section{Introduction}

Contemporary times have witnessed a revolution in string phenomenology, the culmination of decades of enterprise
toward the comprehension of a fundamental high energy theory capable of describing the evolution of our observable universe.
An unwavering theme that has emerged from this century of innovation is nature's persistent rejection of an intransigent
conception of the macrocosm, of which we are just a simple element.  Nature's truths have been revealed in pieces and in
paradoxes, and have stymied every effort to claim mastery over her mysteries.  Whether it be relativistic space and time,
quantum entanglement, or black hole event horizons, we have become acclimated to radical revisions in our sense of reality,
recognizing that the course of time may force all to acquiesce to axioms initially seeming exotic and fantastic,
if they be first synthesized upon rigorous physical maxims.

Progress in the understanding of consistent, meta-stable vacua of string, M- or (predominantly) F-theory flux compactifications
has inspired dramatic challenges to the perspective of our prominence in the cosmos.  Case in point, it has been postulated that
a vast landscape of an astonishing $10^{500}$~\cite{Denef:2004dm,Denef:2007pq} vacua can manifest plausible phenomenology in general.
This suggestion implores inquiry as to why our peculiar vacuum transpired out of the landscape.  One prevalent philosophy contends that
any physically existent universe, whether latent or mature, should correspond to an extremization of probability density in the primordial
quantum froth.  Known as the \textit{Anthropic Principle}, this idea implies that our universe, due to its natural existence and presumed
singularity, occupies a statistical zenith.  Consequently though, this doctrine becomes incurably burdened with fine-tuning complications
of the physical properties of our universe. Motivated by the string landscape and other cosmological scenarios, the speculation of a
Multiverse germinated as a strategy for overcoming those obstacles endemic to fine-tuning.

In our contemporary \textit{Multiverse Blueprints}~\cite{Li:2011dw} we advanced an alternate perspective of our cosmological origins.
We suggested that a mere non-zero probability for a universe featuring our measured physical parameters is the necessary and sufficient
condition.  An observer may inhabit a universe bearing simply a probability of existence which is greater than zero, and not inevitably
that which is most probable.  Moreover, we argued for the significance of No-Scale Supergravity as a universal foundation allowing
for the spontaneous quantum emergence of a cosmologically flat universe.  Experimental validation of a No-Scale ${\cal F}$-$SU(5)$ structure
for our own universe at the LHC could thus reinforce the role of string, M- and F-theory as a master theory of the Multiverse, with No-Scale
supergravity providing an essential model building infrastructure.

\newpage

\begin{figure*}[htp]
	\centering
	\includegraphics[width=1.00\textwidth]{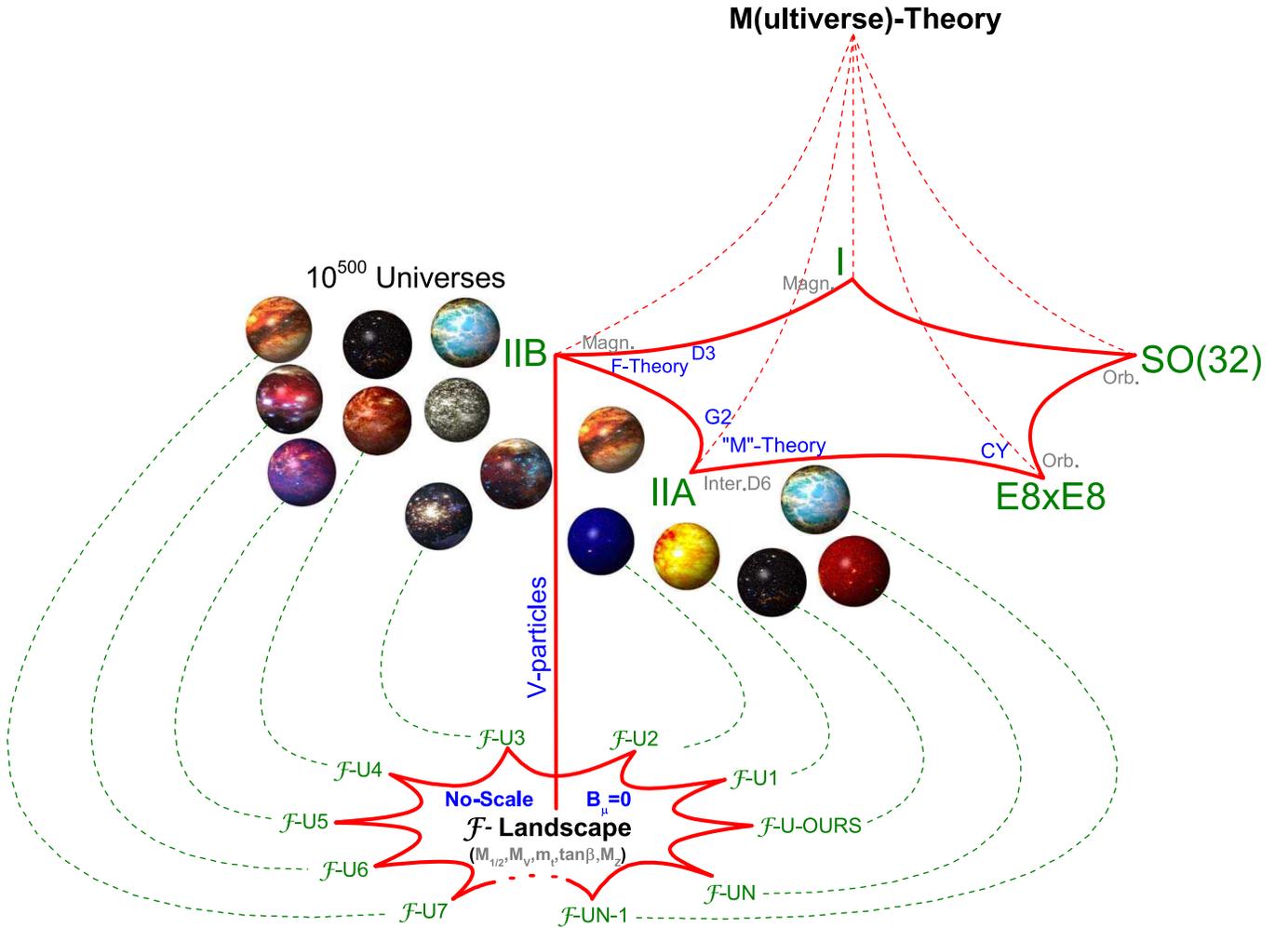}
	\caption{Contemporary perspective on the String Landscape and M-Theory, where we build the M(ultiverse)-Theory with the ${\cal F}$-Landscape derived out of the tripodal foundation in Fig.~\ref{fig:MVII_MultiversePyramid}.}
	\label{fig:MVII_FLandscape_Stars}
\end{figure*}

\begin{figure*}[htp]
	\centering
	\includegraphics[width=1.00\textwidth]{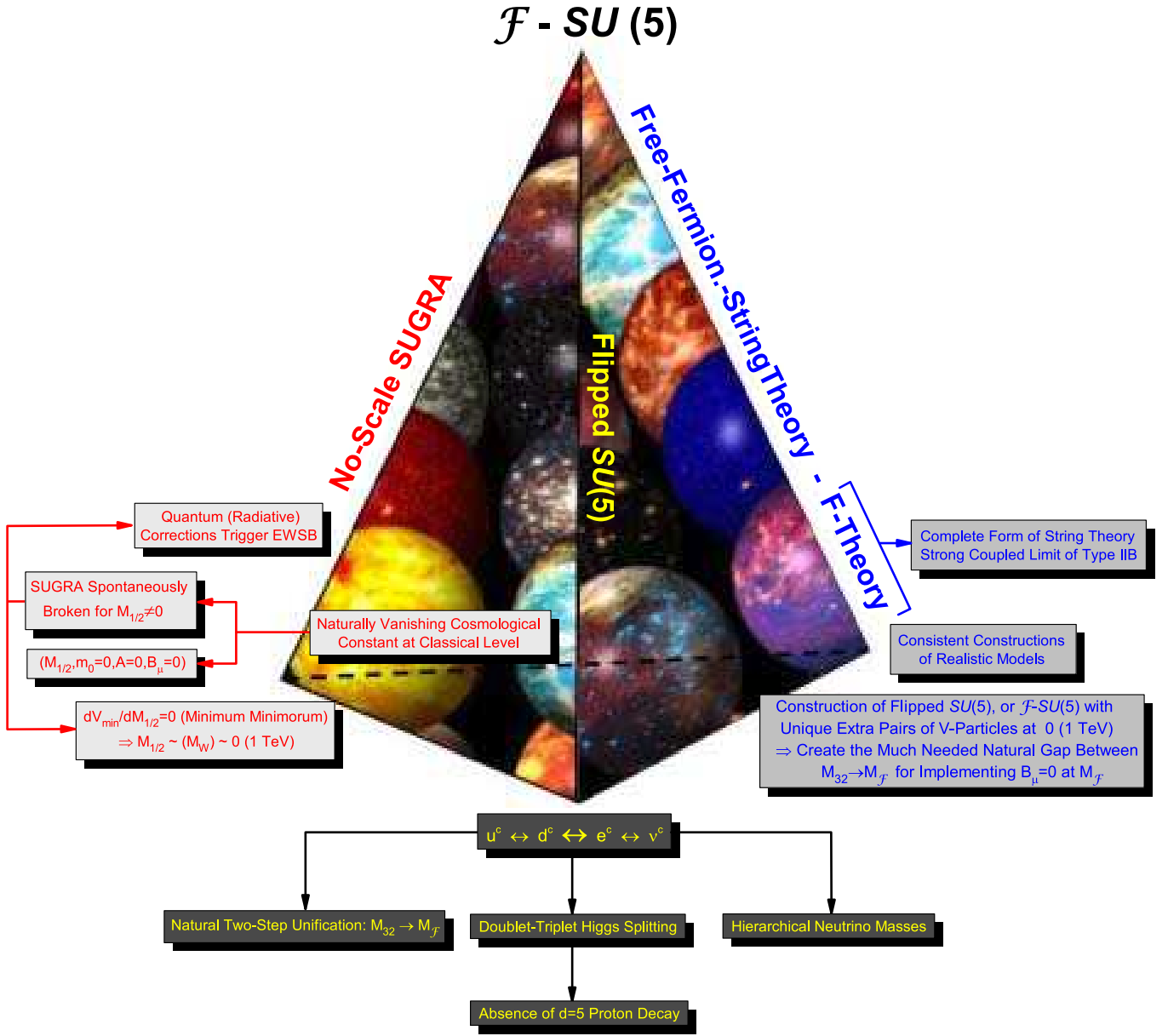}
	\caption{Tripodal foundation of ${\cal F}$-$SU(5)$, built upon the Flipped $SU(5)$ Grand Unified Theory (GUT), extra TeV-Scale vector-like multiplets derived out of F-theory, and the dynamics of No-Scale Supergravity.}
	\label{fig:MVII_MultiversePyramid}
\end{figure*}

We now undertake a first task of engineering in association with our Blueprints~\cite{Li:2011dw}, considering the possibility of Multiverse model building,
or \textit{universe building}.  We employ a precision numerical analysis to derive and subsequently classify the features of the No-Scale ${\cal F}$-$SU(5)$
Multiverse, within some local neighborhood of our own universe's phenomenology.  By secondarily minimizing each model's scalar Higgs potential minima, under
application of the dynamic Super No-Scale condition~\cite{Li:2010uu,Li:2011dw,Li:2011xu}, only legitimate electroweak symmetry breaking (EWSB) vacua are
viable elements of the solution space. The dynamically selected EWSB vacuum at this point of secondary minimization, which is in correspondence with the stabilization
of a string-theoretic modulus, will be identified as the \textit{minimum minimorum} (MM).  Thereupon, \textit{all} MM realize our minimal specifications for a greater
than zero probability of emerging from the landscape.  Hence, we conclude that a contiguous hyperspace of MM in No-Scale ${\cal F}$-$SU(5)$ may fulfill the intended
goal of constructing the set of locally adjacent Multiverse constituents, endogenous to the plausible solution set of M- and F-Theory flux compactifications. We stress
that application of the dynamic MM vacuum selection criterion elevates the conceptual Multiverse design presented here above a mere scan of the parameter space.
We suggest that the resulting construction might rather be regarded to represent a local dominion of independent universes.

Our paper is organized as follows. First, we shall discuss No-Scale ${\cal F}$-$SU(5)$ in M- and F-Theory flux compactifications,
presenting our ${\cal F}$-$SU(5)$ M(ultiverse)-Theory. Next, we engage in a brief review of ${\cal F}$-$SU(5)$, the Super No-Scale condition,
and our secondary minimization procedure. In the latter half of our work, we shall demonstrate the minimization of discrete elements within the model space,
and extrapolate the results to construct a hyperspace of MM, interpreting the solution space in terms of our local community of universes within the Multiverse. We then present phenomenology of a distinctive universe within the larger Multiverse structure that can explain tantalizing experimental hints of the higgs boson and supersymmetry within our own universe.

\section{The ${\cal F}$-$SU(5)$ M(ultiverse)-Theory}

The Standard Model has been confirmed as a correct effective field
theory valid up to about 100~GeV. Nonetheless, problems exist, such as the gauge hierarchy problem,
charge quantization, and an excessive number of parameters, etc. Moreover, the Standard Model excludes gravity. An elegant solution to the gauge hierarchy
problem is supersymmetry. In particular, gauge coupling
unification~\cite{Ellis:1990zq, Ellis:1990wk, Amaldi:1991cn, Langacker:1991an, Anselmo:1991uu, Anselmo:1992jb} can be realized in the supersymmetric SM (SSM), which
strongly implies the Grand Unified Theories (GUTs). In the
GUTs, not only can we explain the charge quantization, but
also reduce the Standard Model parameters due to unification.
Therefore, the interesting question is whether there exists a fundamental
quantum theory or a final theory that can unify the SSM/GUTS 
and general relativity?

The most promising candidate for such a theory is superstring theory.
Superstring theory is anomaly free only in ten dimensions, hence
the extra six space dimensions must be compactified.
As portrayed in Fig.~\ref{fig:MVII_FLandscape_Stars}, there are five consistent ten-dimensional superstring theories:
 heterotic $E_8 \times E_8$, heterotic $SO(32)$, Type I $SO(32)$, Type IIA, 
and Type IIB. Though, this leaves open the question of final unification.
Interestingly, Witten pointed out that this distinction is an artifact of
perturbation theory, and non-perturbatively these five superstring theories
are unified into an eleven-dimensional M-theory~\cite{Witten:1995ex}.
In other words, the five superstring theories are the different perturbative
limits of M-theory. Moreover,  the twelve-dimensional 
F-theory can be considered as the strongly coupled formulation of the Type IIB
string theory with a varying axion-dilaton field~\cite{Vafa:1996xn}, as shown in Fig.~\ref{fig:MVII_FLandscape_Stars}. 

The goal of string phenomenology is to construct the realistic
string vacua, where the SSM/GUTs can be realized and the moduli
fields can be stabilized. Such constructions will give us a bridge
between the string theory and the low energy realistic particle
physics, such that we may test the string models at the Large
Hadron Collider (LHC). Initially, string phenomenology
was studied mainly in the weakly coupled heterotic string theory.
On the other hand, we illustrate in Fig.~\ref{fig:MVII_FLandscape_Stars} that in addition to its perturbative
heterotic string theory corner, M-Theory unification possesses the other corners such as perturbative Type I, Type IIA
and Type IIB superstring theory, which should provide new potentially phenomenologically
interesting four-dimensional string models, related to the heterotic models via 
a web of string
dualities. Most notably, with the advent of D-branes~\cite{Polchinski:1995mt}, 
we can construct the 
phenomenologically interesting string models in Type I, Type
IIA and Type IIB string theories.
Recall that there are five kinds of string models which have 
been studied extensively: (1)  Heterotic $E_8\times E_8$ string model building. The
supersymmetric SM and GUTs can be constructed via
the orbifold 
compactifications~\cite{Buchmuller:2005jr, Lebedev:2006kn, Kim:2006hw} 
and the Calabi-Yau manifold 
compactifications~\cite{Braun:2005ux, Bouchard:2005ag};
(2) Free fermionic string model building. Realistic models
with clean particle spectra can only be constructed at 
the Kac-Moody level one~\cite{Antoniadis:1987tv, 
Antoniadis:1988tt, Antoniadis:1989zy,
 Faraggi:1989ka,
Antoniadis:1990hb, Lopez:1992kg, Cleaver:2001ab}. Note that the Higgs
fields in the adjoint representation or higher can not be
generated at the Kac-Moody level one, so only three kinds
of models can be constructed: the Standard-like models,
Pati-Salam models, and flipped $SU(5)$ 
models~\cite{Antoniadis:1987tv, 
Antoniadis:1988tt, Antoniadis:1989zy,
 Faraggi:1989ka,
Antoniadis:1990hb, Lopez:1992kg, Cleaver:2001ab}.
(3) D-brane model building from Type I, Type IIA,
and Type IIB theories. There are two major kinds of such
models: (i) Intersecting D-brane models or magnetized
D-brane models~\cite{Berkooz:1996km,
Ibanez:2001nd, Blumenhagen:2001te, Cvetic:2001tj,
Cvetic:2001nr, Cvetic:2002pj, Cvetic:2004ui, Cvetic:2004nk,
 Chen:2005aba, Chen:2005mm, Chen:2005mj, Blumenhagen:2005mu};
(ii) Orientifolds of Gepner 
models~\cite{Dijkstra:2004ym, Dijkstra:2004cc}. 
(4) M-theory on $G_2$ manifolds~\cite{Acharya:2001gy, Friedmann:2002ty}. 
Those models can be dual
to the heterotic models on Calabi-Yau threefolds or
to some Type II orientifold models.
(5) F-theory GUTs~\cite{Beasley:2008dc, Beasley:2008kw, Donagi:2008ca, Donagi:2008kj, Jiang:2009zza, Jiang:2009za}. The $SU(5)$ gauge symmetry can be broken
down to the SM gauge symmetries by turning on the $U(1)_Y$
fluxes, and the $SO(10)$ gauge symmetry can be broken down
to the flipped $SU(5)\times U(1)_X$ gauge symmetries and
the $SU(3)_C\times SU(2)_L \times SU(2)_R \times U(1)_{B-L}$
gauge symmetries by turning on the $U(1)_X$ and $U(1)_{B-L}$
fluxes respectively.

To stabilize the moduli fields, the string theories with flux compactifications
have also been studied~\cite{Bousso:2000xa, Giddings:2001yu, Kachru:2003aw, Susskind:2003kw, Denef:2004ze, Denef:2004cf}, in which there intriguingly exist huge meta-stable flux vacua. For example, in the Type IIB theory with
RR and NSNS flux compactifications, the number of the meta-stable 
flux vacua can be of order $10^{500}$~\cite{Denef:2004dm,Denef:2007pq}.
With a weak anthropic principle, this may provide a solution to
the cosmological constant problem and could explain the gauge hierarchy
problem as well.

For our work here in this paper, we study only the flipped $SU(5)\times U(1)_X$ models, and we now shall provide a brief review
of the minimal flipped $SU(5)\times U(1)_X$ model~\cite{Barr:1981qv, Derendinger:1983aj, Antoniadis:1987dx}. 
The gauge group of the flipped $SU(5)$ model is
$SU(5)\times U(1)_{X}$, which can be embedded into $SO(10)$.
We define the generator $U(1)_{Y'}$ in $SU(5)$ as 
\begin{eqnarray} 
T_{\rm U(1)_{Y'}}={\rm diag} \left(-\frac{1}{3}, -\frac{1}{3}, -\frac{1}{3},
 \frac{1}{2},  \frac{1}{2} \right).
\label{u1yp}
\end{eqnarray}
The hypercharge is given by
\begin{eqnarray}
Q_{Y} = \frac{1}{5} \left( Q_{X}-Q_{Y'} \right).
\label{ycharge}
\end{eqnarray}
In addition, 
there are three families of SM fermions 
whose quantum numbers under the $SU(5)\times U(1)_{X}$ gauge group are
\begin{eqnarray}
F_i={\mathbf{(10, 1)}},~ {\bar f}_i={\mathbf{(\bar 5, -3)}},~
{\bar l}_i={\mathbf{(1, 5)}},
\label{smfermions}
\end{eqnarray}
where $i=1, 2, 3$. 

To break the GUT and electroweak gauge symmetries, we 
introduce two pairs of Higgs fields
\begin{eqnarray}
&H={\mathbf{(10, 1)}},~{\overline{H}}={\mathbf{({\overline{10}}, -1)}},& \\ \nonumber
&~h={\mathbf{(5, -2)}},~{\overline h}={\mathbf{({\bar {5}}, 2)}}.&
\label{Higgse1}
\end{eqnarray}
Interestingly, we can naturally solve the doublet-triplet splitting
 problem via the missing partner mechanism~\cite{Antoniadis:1987dx}, and then
the dimension five
proton decay from the colored Higgsino exchange can be
highly suppressed~\cite{Antoniadis:1987dx}.
The flipped $SU(5)\times U(1)_X$ models have been
constructed systematically in the free fermionic string 
constructions at Kac-Moody level one previously~\cite{Antoniadis:1987dx,Antoniadis:1987tv, Antoniadis:1988tt, Antoniadis:1989zy,Lopez:1992kg},
and in the  F-theory model building recently~\cite{Beasley:2008dc, Beasley:2008kw, Donagi:2008ca, Donagi:2008kj, Jiang:2009zza, Jiang:2009za}, and we represent the flipped $SU(5)\times U(1)_X$ models as one pillar of the foundation for ${\cal F}$-$SU(5)$ in Fig.~\ref{fig:MVII_MultiversePyramid}.
In the flipped $SU(5)\times U(1)_X$ models, there are two unification
scales: the $SU(3)_C\times SU(2)_L$ unification scale $M_{32}$ and
the $SU(5)\times U(1)_X$ unification scale $M_{\cal F}$.
To separate the $M_{32}$ and $M_{\cal F}$ scales
and obtain true string-scale gauge coupling unification in 
free fermionic string models~\cite{Jiang:2006hf, Lopez:1992kg} or
the decoupling scenario in F-theory models~\cite{Jiang:2009zza, Jiang:2009za},
we introduce vector-like particles which form complete
flipped $SU(5)\times U(1)_X$ multiplets, and we insert the vector particles and F-Theory as a second pillar in Fig.~\ref{fig:MVII_MultiversePyramid}, and also integrate their presence into Fig.~\ref{fig:MVII_FLandscape_Stars}.
In order to avoid the Landau pole
problem for the strong coupling constant, we can only introduce the
following two sets of vector-like particles around the TeV 
scale~\cite{Jiang:2006hf}
\begin{eqnarray}
&& Z1:  XF ={\mathbf{(10, 1)}}~,~
{\overline{XF}}={\mathbf{({\overline{10}}, -1)}}~;~\\
&& Z2: XF~,~{\overline{XF}}~,~Xl={\mathbf{(1, -5)}}~,~
{\overline{Xl}}={\mathbf{(1, 5)}}
~,~\,
\end{eqnarray}
where 
\begin{eqnarray}
{XF} ~\equiv~ (XQ,XD^c,XN^c)~,~~~{\overline{Xl}}_{\mathbf{(1, 5)}}\equiv XE^c ~.~\,
\end{eqnarray}
In the prior, $XQ$, $XD^c$, $XE^c$, $XN^c$ have the same quantum numbers as the
quark doublet, the right-handed down-type quark, charged lepton, and
neutrino, respectively. 
Such kind of the models have been constructed 
systematically in the F-theory model building locally and dubbed 
${\cal F}-SU(5)$ within that context~\cite{Jiang:2009zza, Jiang:2009za}.
In this paper, we only consider the flipped
$SU(5)\times U(1)_X$ models with 
$Z2$ set of vector-like particles.
The discussions for the models with 
$Z1$ set and heavy threshold corrections~\cite{Jiang:2009zza, Jiang:2009za}
are similar.

Recently, both ATLAS and CMS Collaborations announced the suggestive events
for the Higgs particle with mass around 125 GeV, with each around $2 \sigma$ 
significance over background~\cite{PAS-HIG-11-032, ATLAS-CONF-163}. However,
careful numerical analysis of the viable No-Scale \fsu5 parameter space
yields a prediction for $m_h$ in the range of 119.0~GeV to 123.5~GeV~\cite{Li:2011xg},
consistent with limits from the CMS~\cite{PAS-HIG-11-022}, ATLAS~\cite{ATLAS-CONF-135,ATLAS:2011ww},
CDF and D\O~Collaborations~\cite{:2011ra}.  To increase the
lightest CP-even Higgs boson mass, we consider the Yukawa interaction terms between 
the MSSM Higgs and the vector-like particles in the superpotential
\begin{equation}
W = {\frac{1}{2}} Y_{xd} \, XF \, XF \, h + {\frac{1}{2}} Y_{xu} \, \overline{XF} \, \overline{XF} \, \overline{h}
\end{equation}
After the $SU(5)\times U(1)_X$ gauge symmetry is broken down
to the SM, the relevant Yukawa couplings are
\begin{equation}
W =  Y_{xd} XQ XD^c H_d + Y_{xu} XQ^c XD H_u~.
\end{equation}
Interestingly, we can increase the lightest CP-even Higgs boson mass by around
3-4 GeV via quantum corrections from vector-like particle Yukawa couplings.

\section{Super No-Scale Supergravity}

We now turn to the third and final pillar of the ${\cal F}$-$SU(5)$ foundation in Fig.~\ref{fig:MVII_MultiversePyramid}, that of No-Scale supergravity. In the traditional framework, 
supersymmetry is broken in 
the hidden sector, and then its breaking effects are
mediated to the observable sector
via gravity or gauge interactions. In GUTs with
gravity mediated supersymmetry breaking, also known as the
minimal supergravity (mSUGRA) model, 
the supersymmetry breaking soft terms can be parameterized
by four universal parameters: the gaugino mass $M_{1/2}$,
scalar mass $M_0$, trilinear soft term $A$, and
the ratio of Higgs VEVs $\tan \beta$ at low energy,
plus the sign of the Higgs bilinear mass term $\mu$.
The $\mu$ term and its bilinear 
soft term $B_{\mu}$ are determined
by the $Z$-boson mass $M_Z$ and $\tan \beta$ after
the electroweak (EW) symmetry breaking.

To solve the cosmological constant
problem, No-Scale supergravity was proposed~\cite{Cremmer:1983bf,Ellis:1983sf, Ellis:1983ei, Ellis:1984bm, Lahanas:1986uc}. 
No-scale supergravity is defined as the subset of supergravity models
which satisfy the following three constraints~\cite{Cremmer:1983bf,Ellis:1983sf, Ellis:1983ei, Ellis:1984bm, Lahanas:1986uc}:
(i) The vacuum energy vanishes automatically due to the suitable
 K\"ahler potential; (ii) At the minimum of the scalar
potential, there are flat directions which leave the 
gravitino mass $M_{3/2}$ undetermined; (iii) The super-trace
quantity ${\rm Str} {\cal M}^2$ is zero at the minimum. Without this,
the large one-loop corrections would force $M_{3/2}$ to be either
zero or of Planck scale. A simple K\"ahler potential which
satisfies the first two conditions is
\begin{eqnarray} 
K &=& -3 {\rm ln}( T+\overline{T}-\sum_i \overline{\Phi}_i
\Phi_i)~,~
\label{NS-Kahler}
\end{eqnarray}
where $T$ is a modulus field and $\Phi_i$ are matter fields.
The third condition is model dependent and can always be satisfied in
principle~\cite{Ferrara:1994kg}. We emphasize that No-Scale
supergravity can be realized in the compactification
of the weakly coupled heterotic string 
theory~\cite{Witten:1985xb} and the
compactification of M-theory on S1/Z2 at the
 leading order~\cite{Li:1997sk}.

The scalar fields in the above
K\"ahler potential parameterize the coset space
$SU(N_C+1, 1)/(SU(N_C+1)\times U(1))$, where $N_C$ is the number
of matter fields. Analogous structures appear in the 
$N\ge 5$ extended supergravity theories~\cite{Cremmer:1979up}, for example,
$N_C=4$ for $N=5$, which can be realized in the compactifications
of string theory~\cite{Witten:1985xb, Li:1997sk}. 
The non-compact structure of the symmetry
implies that the potential is not only constant but actually
identical to zero. In fact, one can easily check that
the scalar potential is automatically positive semi-definite,
and has a flat direction along the $T$ field. It is interesting that
for the simple K\"ahler potential in Equation~(\ref{NS-Kahler}),
we obtain the simplest No-Scale boundary condition
$M_0=A=B_{\mu}=0$, while $M_{1/2}$ may be
non-zero at the unification scale,
allowing for low energy SUSY breaking. For an early attempt along these lines, see~\cite{superworld}.

The single relevant modulus field in the simplest 
string No-Scale supergravity is the K\"ahler
modulus $T$, a characteristic of the Calabi-Yau manifold,
the dilaton coupling being irrelevant.
The F-term of $T$ generates the gravitino mass $M_{3/2}$, 
which is proportionally equivalent to $M_{1/2}$.
Exploiting the simplest No-Scale boundary condition at $M_{\cal F}$ and 
running from high energy to low energy under the RGEs,
there can be a secondary minimization, or MM, of the minimum of the
Higgs potential $V_{\rm min}$ for the EWSB vacuum.
Since $V_{\rm min}$ depends on $M_{1/2}$, the gaugino mass $M_{1/2}$ is consequently 
dynamically determined by the equation $dV_{\rm min}/dM_{1/2}=0$,
aptly referred to as the Super No-Scale mechanism~\cite{Li:2010uu,Li:2011dw,Li:2011xu}. In this paper, we shall define the universe as the MM of the effective Higgs potential for a given set of input parameters.

\section{The ${\cal F}$-Landscape}

The methodology of Refs.~\cite{Li:2011dw,Li:2011xu} for computing the MM  had only been applied to a single point within the viable
${\cal F}$-$SU(5)$ parameter space in our previous work. We now seek to originate a full \textit{landscape} of the local ${\cal F}$-$SU(5)$ model space by calculating
the MM for a discrete set of points representative of the neighboring model space that presently subsists in the vicinity of the experimental uncertainties of our own universe~\cite{Li:2011ex}.
Subsequently, we extrapolate the sampled findings to estimate a hypervolume of solutions for a more comprehensive panorama of the model space.  We shall then interpret this
landscape in the context of the \textit{Multiverse Blueprints}~\cite{Li:2011dw}, designating this subdivision as our local Multiverse community.  In a broader sense, the
Multiverse landscape is, of course, not limited to that zone which lies within our experimental uncertainty, though our purpose here is only initially to seek the prospective
structure of an ${\cal F}$-$SU(5)$ local Multiverse, within an acceptable introductory level of precision.  Each point within this No-Scale ${\cal F}$-$SU(5)$ Multiverse
landscape of solutions, which we shall heretofore refer to as the ${\cal F}$-Landscape, can be interpreted as a distinct universe within our regional dominion of universes.  With application of rigorous numerics, we shall demonstrate the resulting solution space. As elaborated in~\cite{Li:2011dw}, testing of the
No-Scale ${\cal F}$-$SU(5)$ framework at the LHC is in some sense likewise a broader test of the framework of the String Landscape and the Multiverse of plausible string,
M- and F-theory vacua.  One can boldly speculate that substantiation of a No-Scale ${\cal F}$-$SU(5)$ configuration for our universe at the LHC offers indirect support for
a local dominion of ${\cal F}$-$SU(5)$ universes.

\begin{figure*}[htp]
        \centering
        \includegraphics[width=0.90\textwidth]{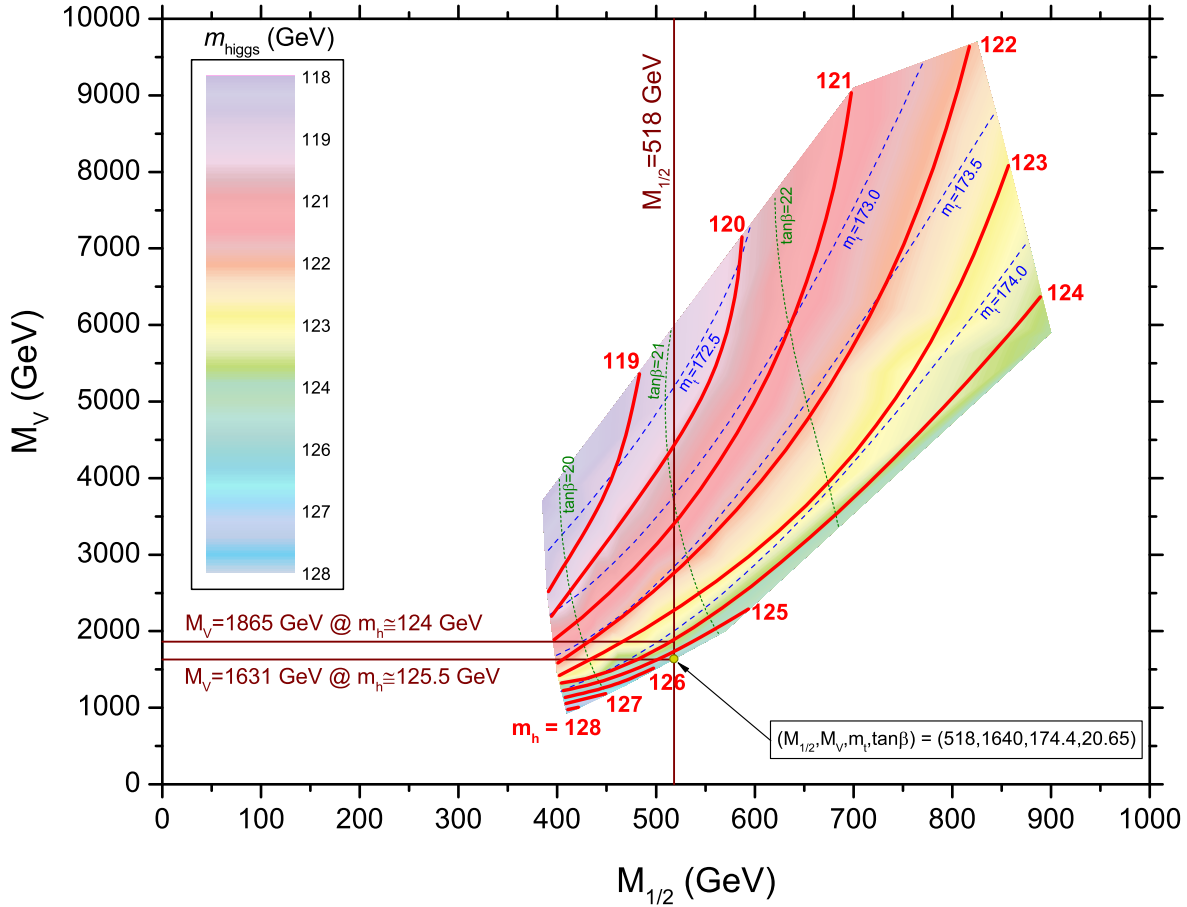}
        \caption{The space of bare-minimal constraints~\cite{Li:2011xu} on the No-Scale \fsu5 model is presented
	in the ($M_{1/2}$,$M_{\rm V}$) plane, with contour overlays designating the $\tan \beta$ and $m_{\rm t}$
	parameter ranges, in addition to the corrected Higgs mass $m_h$, inclusive of the shift from vector-like multiplet interactions.}
        \label{fig:Higgs_wedge}
\end{figure*}

A modest sampling of satisfactory ${\cal F}$-$SU(5)$ points are extracted from the experimentally viable parameter space that satisfies the ``bare-minimal'' constraints
of~\cite{Li:2011xu}, in order to compute the MM in conformity with our conventional methodology. With the vector-like particle
contributions to the Higgs boson mass, we present the updated viable parameter space in Fig.~(\ref{fig:Higgs_wedge}).
The outermost borders of the experimentally
viable parameter space presented in~\cite{Li:2011xu} are circumscribed from the bare-minimal constraints, though these constraints in principle are applicable only to our
universe and not the Multiverse in general. Nevertheless, the model space persisting within this constrained perimeter presents a generous supply of archetype universes to
explore and accordingly construct a hypervolume of solutions. To recapitulate, the bare-minimal constraints for our universe are defined by compatibility with the world
average top quark mass $m_{\rm t}$ = $173.3\pm 1.1$ GeV~\cite{:1900yx}, the prediction of a suitable candidate source of cold dark matter (CDM) relic density matching the upper and lower thresholds $0.1088 \leq \Omega_{CDM} \leq 0.1158$ set by the WMAP-7 measurements~\cite{Komatsu:2010fb}, a rigid prohibition against a charged lightest supersymmetric particle (LSP), compatibility with the precision LEP constraints on the lightest CP-even Higgs boson ($m_{h} \geq 114$ GeV~\cite{Barate:2003sz,Yao:2006px}) and other light SUSY chargino, stau, and neutralino mass content, and a self-consistency specification on the dynamically evolved value of $B_\mu$ measured at the boundary scale $M_{\cal{F}}$. An uncertainty of $\pm 1$~GeV on $B_\mu = 0$ is allowed, consistent with the induced variation from fluctuation of the strong coupling within its error bounds and the expected scale of radiative electroweak (EW) corrections. The lone constraint above that is necessarily mandatory for the Multiverse is that of the condition on the B-parameter at the $M_{\cal{F}}$ scale, since there is certainly no prerequisite for any of these other constrained parameters to inhabit within or even adjacent to the experimentally established uncertainties for our universe, although for our study here we prefer to remain nearby the local experimental ambiguities. The cumulative effect of these bare-minimal constraints distinctively shapes the experimentally viable parameter space germane to our universe into the uniquely formed profile situated in the ($M_{1/2},M_{\rm V}$) plane exhibited in Fig.~(\ref{fig:Higgs_wedge}), from a tapered light mass region with a lower bound of $\tan \beta$ = 19.4 into a more expansive heavier region that ceases sharply with the charged stau LSP exclusion around tan$\beta \simeq$ 23. Correspondingly, we shall not journey too far afield from this narrow region of tan$\beta$ or the world average top quark periphery.

\begin{figure*}[htp]
	\centering
	\includegraphics[width=0.95\textwidth]{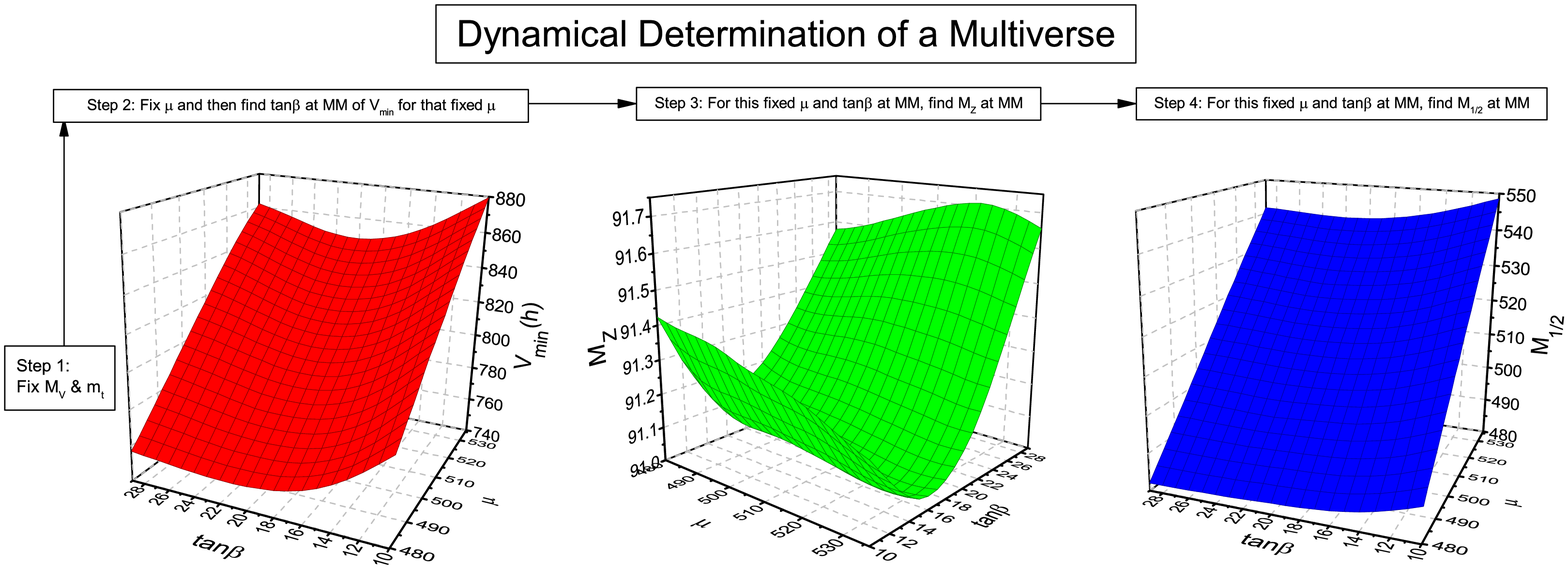}
	\includegraphics[width=0.95\textwidth]{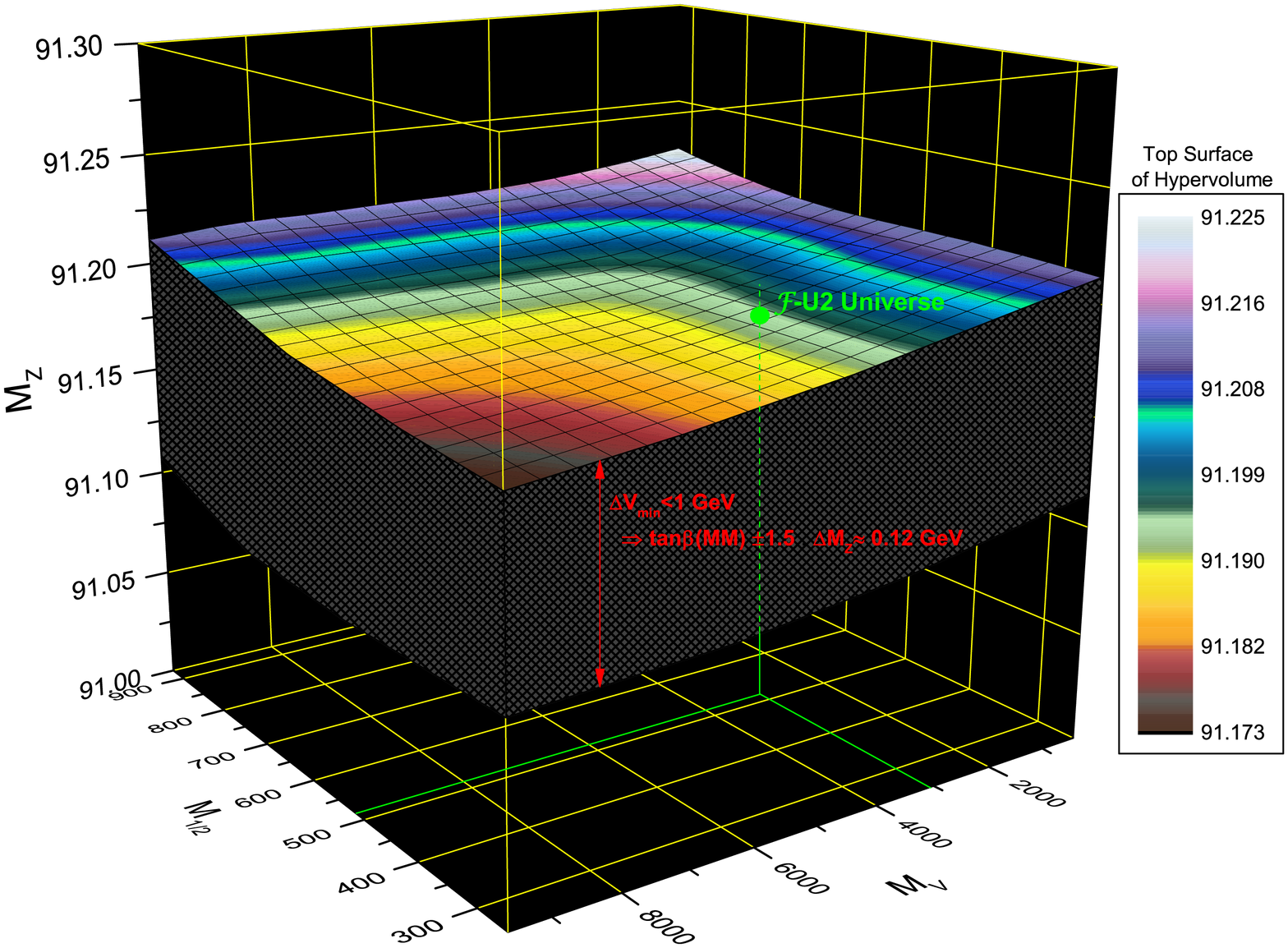}
	\caption{Logical flow depicting the process of dynamically determining a Multiverse. The upper three plot spaces show the $B_{\mu}$=0 hypersurfaces for a fixed set of ($M_V,m_t$). The lower plot space is generated by dynamically determining numerous points throughout the model space to estimate the hypervolume of minimum minimorum. Here we show the coordinates of a benchmark
universe ${\cal F}$-U2, which represents a plausible candidate for our universe. The thickness of the volume is approximated by placing a constraint of $\Delta V_{min}(h) <$ 1 GeV at the minimum minimorum, similar in scale to the QCD corrections at the second loop. This results in a deviation of $\pm$1.5 on the tan$\beta$ at the minimum minimorum, translating into a 0.12 GeV uncertainty on the dynamically determined value of the electroweak scale $M_Z$.}
	\label{fig:MVII_Flowchart}
\end{figure*}

The production of the hypervolume of solutions is initiated by mining the bare-minimally constrained wedge region in Fig.~(\ref{fig:Higgs_wedge}) for prospective universes from which to compute $V_{min}(h)$, carrying precision equivalent to the LEP constraints on the electroweak scale $M_Z$. In~\cite{Li:2011dw,Li:2011xu}, we executed the minimization procedure for a single specific fixed numerical value of $\mu$ only, so in essence, here we are broadening the blueprint of~\cite{Li:2011dw,Li:2011xu} to encompass an extensive range of $\mu$, utilizing our prescribed freedom of the numerical parameter. The secondary minimization procedure is thus enlarged by an order of cardinality, such that we may position the numerical value of $\mu$ to any figure we require, essentially dynamically determining in principle all $M_{1/2}$, tan$\beta$, and $M_{Z}$ for any preset permutation of $M_{V}$ and $m_{t}$.  This prescription can be replicated for an indefinite quantity of regional points within the model space in order to extrapolate the outcome to an estimated hypervolume comprising our local dominion of universes.  A logically sequenced rendering of the prescription for dynamically determining a Multiverse is illustrated in Fig.~\ref{fig:MVII_Flowchart}, with the top half of the Fig.~\ref{fig:MVII_Flowchart} space elucidating the minimization procedure for a unique predetermined duo of $M_{V}$ and $m_{t}$, while the bottom half of the plot space reveals a depiction of the conjectural hypervolume of universes.

Explicitly, the flow demonstrated in Fig.~\ref{fig:MVII_Flowchart}, after selection of a fixed combination of $M_{V}$ and $m_{t}$, proceeds first to pinpoint tan$\beta$ at the minimum of the 1-loop Higgs potential $V_{min}(h)$ for a precise numerical value of $\mu$, as depicted in the upper left element, which is now deemed the MM. The curved grid surfaces illustrated in the top half of the Fig.~\ref{fig:MVII_Flowchart} space characterize the hypersurface of $B_\mu = 0$ solutions. The effect of the $\pm 1$~GeV induced electroweak scale variations on the $B_\mu = 0$ condition translates into a small thickness of the $B_\mu = 0$ hypersurfaces in the top half of the Fig.~\ref{fig:MVII_Flowchart} space, though we suppress this in the diagrams here for simplicity. At first glance, tan$\beta$ at the MM appears to be constant in Fig.~\ref{fig:MVII_Flowchart}, though in fact it is not, as tan$\beta$ at the MM experiences a slight gradual continuous variation as the numerical value of $\mu$ is continuously adjusted. Once tan$\beta$ at the MM for our selection of $\mu$ is discovered, we can then resolve the corresponding $M_{Z}$ and $M_{1/2}$ at this MM by analyzing the center and right plots in the top half of the Fig.~\ref{fig:MVII_Flowchart} space. We have in no way up to this point deviated from the methodology of Refs.~\cite{Li:2011dw,Li:2011xu}. We have only demonstrated that guidelines established in Refs.~\cite{Li:2011dw,Li:2011xu} can be broadened to incorporate the selection of any $\mu$, such that the freedom on the bilinear $\mu$ parameter can in some sense be envisioned as a dial that can ``tune'' $M_{1/2}$, tan$\beta$, and $M_{Z}$ to that of any distinctive universe, for any and all prescribed sets of $M_{V}$ and $m_{t}$, traversing the $B_\mu = 0$ hypersurfaces.

The multistep minimization procedure is copied for a sizable quantity of points in the model space, generating the solution space in the lower half plot of Fig.~\ref{fig:MVII_Flowchart} through an extrapolation of the discrete returns. Only those sub one GeV perturbations about the minimum of the 1-loop Higgs potential are preserved, which we judge to be comparable in scale to the QCD corrections to the Higgs potential at the second loop. This constraint confines the value of tan$\beta$ at the MM to live within an expected $\pm$1.5 deviation around the absolute minimum of $V_{min}(h)$. Consequently, we can project the ensuing variation in $M_{Z}$ to be about $\pm$0.12 GeV at the MM. Thusly, over and above the freedom in $\mu$ to select different universes by ``tuning'' $M_{1/2}$, tan$\beta$, and $M_{Z}$ along a continuous string of MM, we must further recognize the indeterminate nature of these parameters at the MM from the QCD fluctuations providing some discretion on confinement of the MM to this theoretic one-dimensional string. Yet, it is essential to bear in mind that altering any one of these parameters will demand a compensating adjustment in one or more of the remaining parameters in order to transit along the $B_\mu = 0$ direction, engendering an additional unique point in the hypervolume of solutions, i.e. a unique universe in the Multiverse. These small fluctuations about the MM induce the diagrammed thickness of the hypervolume advertised in Fig.~\ref{fig:MVII_Flowchart}, where each singular point in the illustrated hypervolume exemplifies an individual universe in the Multiverse.

The points employed in the compilation of the $B_\mu = 0$ hypersurface and hypervolume of Multiverse solutions in Fig.~\ref{fig:MVII_Flowchart} were extracted from the experimentally viable parameter space delineated in Ref.~\cite{Li:2011xu}, where the contours of tan$\beta$ defining those regions consistent with the WMAP-7 relic density measurements progressively scale with both $M_{1/2}$ and $M_{V}$. As noted earlier, the WMAP-7 experimentally allowed parameter space spans from $\tan \beta$ = 19.4 to around tan$\beta \simeq$ 23, enveloping those regions of the model space regarded as credible contenders for our universe from a bottom-up experimental perspective. From a Multiverse frame of reference, the WMAP-7 region is extraneous, as any universe within the ${\cal F}$-Landscape may possess an intrinsic ``WMAP'' dark matter density, so to speak. In the process of dynamically determining the $M_{1/2}$, tan$\beta$, and $M_{Z}$ at the MM, relevant to the top-down theoretical perspective, there is little reason to anticipate (at least not from the point of view of an island universe) that the bottom-up and top-down techniques should be self-consistent at more than just a single point.
Nevertheless, this remarkable correspondence is unquestionably what is discovered, prompting curiosity at whether the correlation stems from a deep physical motivation.  In particular, the parallel transport of parameterization freedom exhibited by the phenomenological and dynamical treatments appears to support the conjectural application of this framework to a continuum of locally adjacent universes, each individually seated at its own dynamic MM.



\section{LHC Search}

\begin{table}[ht]
  \small
    \centering
    \caption{Spectrum (in GeV) for $M_{1/2} = 518$~ GeV, $M_{V} = 1640$~GeV, $m_{t} = 174.4$~GeV, tan$\beta$ = 20.65. Here, $\Omega_{\chi}$ = 0.1155 and the lightest neutralino is 99.9\% Bino.
	The partial lifetime for proton decay in the leading ${(e|\mu)}^{+} \pi^0 $ channels falls around $4 \times 10^{34}$~Y~\cite{Li:2010dp,Li:2010rz}.}
		\begin{tabular}{|c|c||c|c||c|c||c|c||c|c||c|c|} \hline
    $\widetilde{\chi}_{1}^{0}$&$99$&$\widetilde{\chi}_{1}^{\pm}$&$216$&$\widetilde{e}_{R}$&$196$&$\widetilde{t}_{1}$&$558$&$\widetilde{u}_{R}$&$1053$&$m_{h}$&$125.4$\\ \hline
    $\widetilde{\chi}_{2}^{0}$&$216$&$\widetilde{\chi}_{2}^{\pm}$&$900$&$\widetilde{e}_{L}$&$570$&$\widetilde{t}_{2}$&$982$&$\widetilde{u}_{L}$&$1144$&$m_{A,H}$&$972$\\ \hline
    $\widetilde{\chi}_{3}^{0}$&$896$&$\widetilde{\nu}_{e/\mu}$&$565$&$\widetilde{\tau}_{1}$&$108$&$\widetilde{b}_{1}$&$934$&$\widetilde{d}_{R}$&$1094$&$m_{H^{\pm}}$&$976$\\ \hline
    $\widetilde{\chi}_{4}^{0}$&$899$&$\widetilde{\nu}_{\tau}$&$551$&$\widetilde{\tau}_{2}$&$560$&$\widetilde{b}_{2}$&$1046$&$\widetilde{d}_{L}$&$1147$&$\widetilde{g}$&$704$\\ \hline
		\end{tabular}
		\label{tab:masses}
\end{table}

\begin{figure*}[htp]
        \centering
        \includegraphics[width=0.86\textwidth]{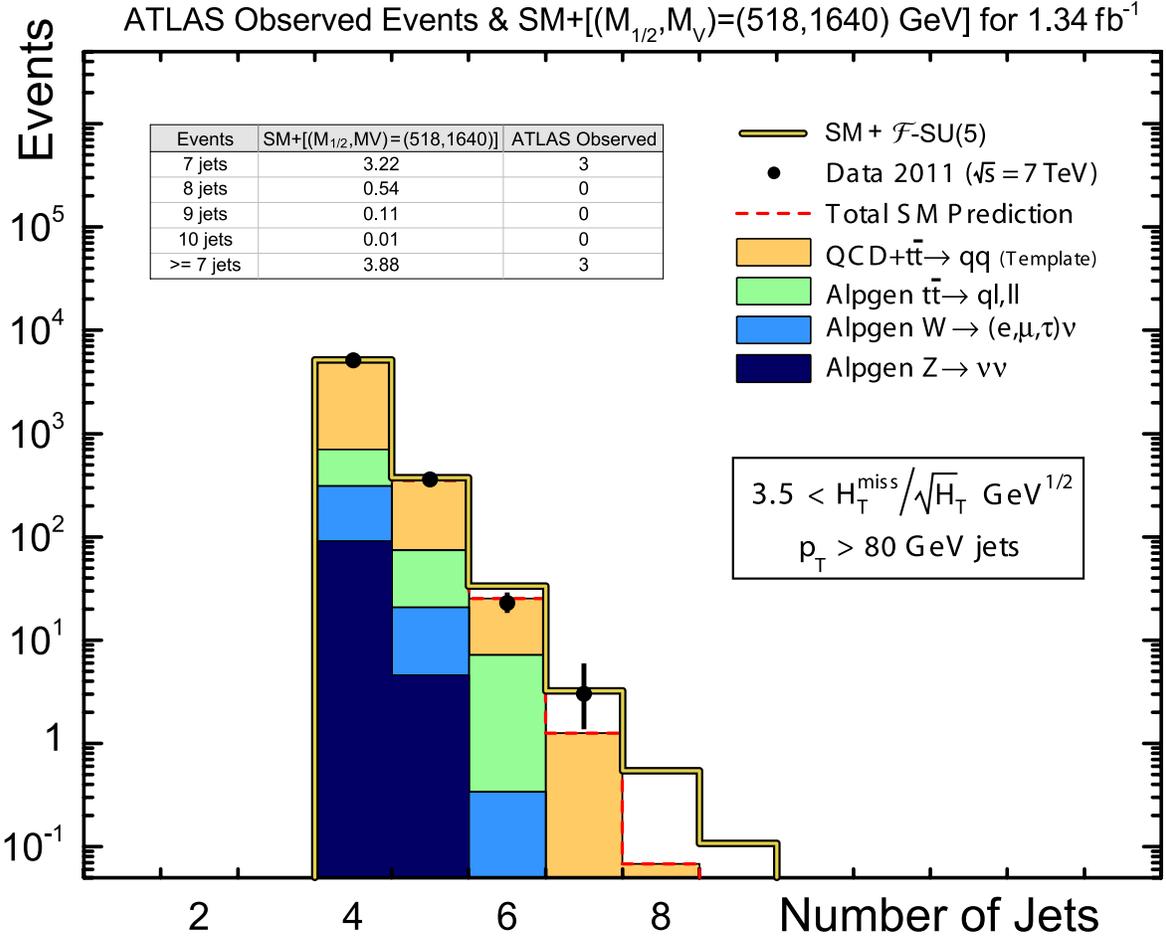}
        \caption{An ATLAS collaboration plot~\cite{Aad:2011qa} (present in the arXiv source repository supplementing
	the cited document) representing $1.34~{\rm fb}^{-1}$ of integrated luminosity
	at $\sqrt{s} = 7$~TeV is reprinted with an overlay summing our Monte Carlo collider-detector simulation of the
	No-Scale \fsu5 model benchmark ($M_{1/2}=518$~GeV, $M_{\rm V}=1640$~GeV) with the ATLAS SM background.}
        \label{fig:ATLAS_data}
\end{figure*}

\begin{figure*}[htp]
        \centering
        \includegraphics[width=0.79\textwidth]{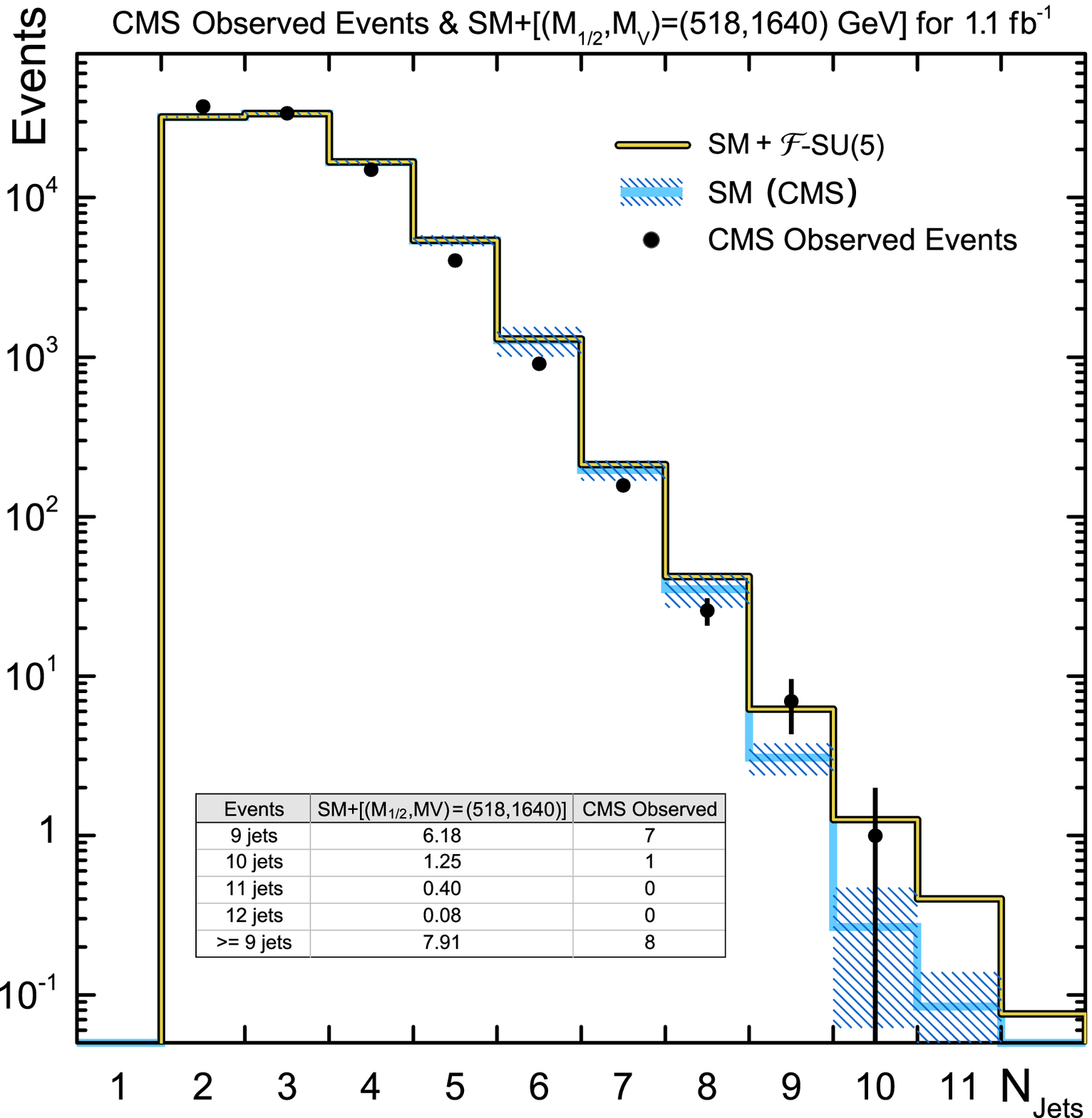}
        \caption{A CMS collaboration plot~\cite{PAS-SUS-11-003} representing $1.1~{\rm fb}^{-1}$ of integrated luminosity
	at $\sqrt{s} = 7$~TeV is reprinted with an overlay summing our Monte Carlo collider-detector simulation of the
	No-Scale \fsu5 model benchmark ($M_{1/2}=518$~GeV, $M_{\rm V}=1640$~GeV) with the CMS SM background.}
        \label{fig:CMS_data}
\end{figure*}

We have carefully studied the expected \fsu5 production excesses in the high multiplicity jet channels~\cite{Maxin:2011hy,Li:2011gh,Li:2011rp,Li:2011fu,Li:2011av},
undertaking a detailed and comprehensive Monte Carlo simulation, employing industry standard tools~\cite{Stelzer:1994ta,MGME,Alwall:2007st,Sjostrand:2006za,PGS4}.
We have painstakingly mimicked~\cite{Maxin:2011hy,Li:2011av} the leading multi-jet selection strategies of the CMS~\cite{PAS-SUS-11-003} and ATLAS~\cite{Aad:2011qa}
collaborations, using a post-processing script of our own design~\cite{cutlhco}.  All 2-body SUSY processes have been included in our simulation.
Our conclusion is that the best fit to the jet production excesses observed at both detectors occurs in the vicinity of the $M_{1/2} = 518$~GeV strip of
Fig.~\ref{fig:Higgs_wedge}.  Lighter values of $M_{1/2}$ will allow for lighter 
vector-like particles
and a heavier top quark, and thus also a heavier Higgs.  However, values much below about $M_{1/2} = 480$ are considered to be
excluded for over-production of SUSY events.  Values much larger than the target range
will have some difficulty achieving a sufficiently large Higgs mass.  
For specificity, we consider a benchmark point with inputs $M_{1/2}=518~{\rm GeV}$,
$M_V=1640$~GeV, $m_{\rm t} = 174.4$ and $\tan\beta=20.65$~\cite{Li:2011ab}.
The SUSY spectrum for this benchmark is presented in Table~(\ref{tab:masses}).
In particular, the lightest CP-even Higgs boson mass can be lifted from 121.4 GeV up
to 125.4 GeV. In Figs.~\ref{fig:ATLAS_data} and \ref{fig:CMS_data}, we overlay counts for
the No-Scale \fsu5 jet production (summed with the official SM backgrounds) onto histograms illustrating the current status of the LHC multi-jet
SUSY search, representing just over $1.1~{\rm fb}^{-1}$ of luminosity integrated by the ATLAS~\cite{Aad:2011qa} and CMS~\cite{PAS-SUS-11-003}
experiments, respectively.  The statistical significance of the ATLAS overproduction, as gauged by the indicator of signal (observations minus background)
to background ratio $S/\sqrt{B+1}$, is quite low for $\ge$ 7 jets in the search strategy of Fig.~(\ref{fig:ATLAS_data}),
somewhat greater than 1.0, although the CMS overproduction significance for $\ge$ 9 jets in the search strategy of Fig.~(\ref{fig:CMS_data}) is just above $2.0$.
We project in Table~(\ref{tab:signals}) that the already collected $5~{\rm fb}^{-1}$ data set may be sufficient to reach the gold standard signal significance
of 5, at least for the CMS search strategy, although both approaches appear to scale well with higher intensities.

\begin{table}[h]
\centering
\caption{Projections for the ATLAS and CMS signal significance at 5~${\rm fb}^{-1}$ of integrated luminosity, in the ultra-high
jet multiplicity channels.  Event counts for \fsu5 are based on our own Monte Carlo of the $M_{1/2} = 518$~GeV, $M_{V} = 1640$~GeV
benchmark.  SM backgrounds are scaled up from official collaboration estimates~\cite{PAS-SUS-11-003,Aad:2011qa}.}
\begin{tabular}{|c|c|c|c|c|c||c|c|c|c|c|}\cline{2-11}
\multicolumn{1}{c|}{} & \multicolumn{5}{|c||}{CMS~$5~{\rm fb}^{-1}$} & \multicolumn{5}{|c|}{ATLAS~$5~{\rm fb}^{-1}$} \\ \cline{2-11}
\multicolumn{1}{c|}{} &$	{\rm 9j}$&$	{\rm 10j}$&$	{\rm 11j}$&$	{\rm 12j}$&$\,{\rm \ge9j}\,$
&$	{\rm 7j}$&$	{\rm 8j}$&$	{\rm 9j}$&$	{\rm 10j}$&$\,{\rm \ge7j}\,$ \\ \hline	\hline
\fsu5 &$	14.0$&$	4.5$&$	1.4$&$	0.3$&$	20.3$&$	7.3$&$	1.8$&$	0.4$&$	0.1$&$	9.6$ \\ \hline
${\rm SM}	$&$	14.0$&$	1.2$&$	0.4$&$	0.0$&$	15.6$&$	4.7$&$	0.3$&$	0.0$&$	0.0$&$	4.9$ \\ \hline \hline
$S/ \sqrt{B+1}$&$	3.6$&$	3.0$&$	1.2$&$	0.3$&${{\bf 5.0}}$&$	3.1$&$	1.6$&$	0.4$&$	0.1$&${{\bf 3.9}}$ \\ \hline
\end{tabular}
\label{tab:signals}
\end{table}





\section{Conclusions}

We evolved our \textit{Multiverse Blueprints} to characterize our local neighborhood of the String Landscape and the Multiverse of
plausible string, M- and F-theory vacua. Considering Super-No-Scale ${\cal F}$-$SU(5)$, we demonstrated the existence of a continuous family of solutions which might adeptly describe
the dynamics of distinctive universes.  This Multiverse landscape of ${\cal F}$-$SU(5)$ solutions, which we  referred to as the
${\cal F}$-Landscape, accommodates a subset of universes compatible with the presently known experimental uncertainties of our own
universe. We showed that by secondarily minimizing the minimum of the scalar Higgs potential of each solution within the ${\cal F}$-Landscape,
a continuous hypervolume of distinct \textit{minimum minimorum} can be engineered which comprise a regional dominion of universes,
with our own universe cast as the bellwether.  In addition, we pointed out that
our model can be tested at the early LHC run, and
 conjectured that an experimental signal at the LHC of the No-Scale ${\cal F}$-$SU(5)$
framework's applicability to our own universe might sensibly be extrapolated as corroborating evidence for the role of string, M- and
F-theory as a master theory of the Multiverse, with No-Scale supergravity as a crucial and pervasive reinforcing structure.


\begin{acknowledgments}
This research was supported in part 
by the DOE grant DE-FG03-95-Er-40917 (TL and DVN),
by the Natural Science Foundation of China 
under grant numbers 10821504 and 11075194 (TL),
by the Mitchell-Heep Chair in High Energy Physics (JAM),
and by the Sam Houston State University
2011 Enhancement Research Grant program (JWW).
We also thank Sam Houston State University
for providing high performance computing resources.
\end{acknowledgments}


\bibliography{bibliography}

\end{document}